\documentclass[manuscript]{aastex}

\usepackage{multirow}
\usepackage{graphicx,color}
\newcommand{\etal }{{et al.} }
\newcommand{\msun}{\thinspace M_\odot}

\newcommand{\rsun}{\thinspace R_\odot} 

\def\lesssim{\mathrel{\hbox{\rlap{\hbox{\lower4pt\hbox{$\sim$}}}\hbox{$<$}}}}
\def\gtrsim{\mathrel{\hbox{\rlap{\hbox{\lower4pt\hbox{$\sim$}}}\hbox{$>$}}}}
\newcommand{\cm}{\,{\rm cm}^{-3} } 
\newcommand{\km}{\,{\rm km\, s}^{-1}}

\newcommand{\mdot}{M_\odot\,{\rm yr}^{-1} }

\newcommand{\dfrac}[2]{{\displaystyle \frac{#1}{#2}} }

\shorttitle{Protostellar Jet}
\shortauthors{Machida}

\begin{document}
\title{Protostellar Jets Enclosed by Low-velocity Outflows}


\author{ Masahiro N. Machida\altaffilmark{1}}
\affil{Department of Earth and Planetary Sciences, Faculty of Sciences, Kyushu University, Fukuoka 812-8581, Japan}

\email{ machida.masahiro.018@m.kyushu-u.ac.jp}


\begin{abstract}
A protostellar jet and outflow are calculated for $\sim270$\,yr following the protostar formation using a three dimensional magnetohydrodynamics simulation, in which both the protostar and its parent cloud are spatially resolved. 
A high-velocity ($\sim100\km$) jet with  good collimation is driven near the disk's inner edge, while a low-velocity ($\lesssim 10\km$) outflow with a wide opening angle appears in the outer-disk region.  
The high-velocity jet propagates into the low-velocity outflow, forming a nested velocity structure in which a narrow high-velocity flow is enclosed by a wide low-velocity flow.
The low-velocity outflow is in a nearly steady state, while the high-velocity jet appears intermittently.
The time-variability of the jet is related to the episodic accretion from the disk onto the protostar, which is caused by gravitational instability and magnetic effects such as magnetic braking and magnetorotational instability.
Although the high-velocity jet has a large kinetic energy, the mass and momentum of the jet are much smaller than those of the low-velocity outflow. 
A large fraction of the infalling gas is ejected  by the low-velocity outflow.
Thus, the low-velocity outflow actually has a more significant effect than the high-velocity jet in the very early phase of the star formation.
\end{abstract}

\section{Introduction} 
Many protostellar jets and outflows have been observed in star forming regions \citep[e.g.,][]{arce07,frank14} and are considered as significant phenomena in the star formation process.
They carry away excess angular momentum and large fractions of gas from star forming clouds into interstellar space and determine both the star formation efficiency and the stellar mass \citep{matzner00,nakano95}. 
Observations indicate that a high-velocity (optical) jet is enclosed by a low-velocity (molecular) outflow \citep{mundt83,arce02,velusamy07,ren11}.
In addition, the high-velocity jets exhibit good collimation, while the low-velocity outflows have wide opening angles. 
Despite their importance, their driving mechanisms and propagation dynamics are poorly understood.

The driving mechanism of outflows and jets has been investigated using numerical simulations. 
The driving is highly dependent on many (ambiguous) parameters, including the velocity and density distributions, the magnetic field  around the protostar, the properties of the infalling envelope and circumstellar disk, and the rate of the accretion onto the disk and protostar  \citep{konigl00,pudritz07}. 
To determine them consistently, it is necessary to calculate the evolution of the star forming cloud from the prestellar cloud stage, resolving both the protostar and its parent cloud.

Using a two-dimensional ideal magnetohydrodynamics (MHD) simulation, \citet{tomisaka98} demonstrated for the first time  the low-velocity outflow driven by the first core in a collapsing cloud. 
His successive studies \citep{tomisaka00,tomisaka02} showed that the low- and high-velocity flows are driven by different objects of the first and second (or protostar) cores. 
\citet{machida08a} also reproduced the high-velocity jet and  low-velocity outflow with their three-dimensional resistive MHD simulations.
Recently, the high- and low-velocity flows driven in the collapsing cloud were confirmed by radiation MHD simulations \citep{tomida13,bate14}. 
However, these calculations were executed only for the first few years following the protostar formation, during which the jet reaches just $\sim1$\,AU from the protostar.  
Thus, it is very difficult to compare simulations with observations, because at this stage the jet is embedded in a dense gas. 
In this study, starting from a prestellar cloud, I extend the calculation to $\sim270$\,yr after  protostar formation, at which time the jet has reached  $\sim400$\,AU from the protostar.

\section{Model and Numerical Method} 
As the initial state, I consider a prestellar cloud with a Bonner-Ebert density profile that is characterized by a central number density $n_{\rm 0}$ and isothermal temperature $T_{\rm iso}$. 
In this study, $n_{\rm 0}=6\times10^5\cm$  and $T_{\rm iso}=10\,K$ are adopted, and the cloud density is enhanced by a factor of $1.5$ to promote cloud contraction. 
The initial cloud radius corresponds to twice the critical Bonner-Ebert radius of $1.2\times10^4$\,AU and is embedded in an interstellar medium that has a uniform number density of $n=1.1\times10^4\cm$.
The mass of the initial cloud is  $1.0\msun$.
A uniform magnetic field  $B_0 =4.5\times10^{-5}$\,G is imposed, and a rigid rotation  $\Omega_0=1.1\times10^{-13}$\,s$^{-1}$ is adopted inside the initial cloud. 
The rotation axis is parallel to the magnetic field. 
The initial cloud has a ratio of rotational to gravitational energy of $\beta_0=0.02$, and mass-to-flux ratio normalized  by the critical value \citep{mouschovias76} of $\mu=1$. 
Note that since the mass-to-flux ratio is a function of radius, it has $\mu > 1$ inside the cloud \citep{machida11b}.
These initial settings are almost identical to those used in previous studies \citep{machida07,machida08a,machida08b,machida11,machida12,machida13a}.


To calculate the cloud evolution, I adopt the nested grid method, which is composed of three-dimensional nested rectangular grids all of which comprise $64\times 64\times 32$ cells.
Mirror symmetry is imposed across the $z=0$ plane. 
The initial cloud exists in the fifth level of the grid, of which the box size is $L_{5}=2.4\times 10^4$\,AU and the cell width is $h_5=370$\,AU.
The box size of the first level of the grid is $L_1=3.8\times10^5$\,AU.
The finer grid is dynamically generated as the cloud collapses to maintain the Truelove condition \citep{truelove97}.
The finest grid level is $l=21$ and  has $L_{21}=0.35$\,AU and $h_{21}=5.6\times10^{-3}$\,AU.

In this study, resistive MHD equations including self-gravity are solved, in which  the barotropic equation of state $P\propto \rho^\Gamma$ is used instead of solving the energy equation.
The polytropic exponent $\Gamma$ is set as
\begin{equation} 
\Gamma = \left\{
\begin{array}{lll}
 1    & {\rm for} &  n < 2\times10^{10} \cm, \\
 5/3  & {\rm for} & 2\times 10^{10} \cm < n < 3\times10^{13} \cm , \\
 7/5  & {\rm for} & 3\times 10^{13} \cm < n < 10^{15} \cm, \\
 1.1  & {\rm for} & 10^{15} \cm < n < 3\times10^{18} \cm, \\
 2    &  {\rm for} & n > 3\times10^{18} \cm.
\label{eq:gamma}
\end{array}
\right.  
\end{equation}
The equation of state in the range $n<3\times10^{18}\cm$ can approximately reproduce the thermal evolution of a collapsing cloud core \citep{larson69,masunaga00,tomida13}.
On the other hand,  a harder equation of state (larger $\Gamma$) in the range of $n>3\times10^{18}\cm$ can stop the increase in density and significantly alleviates the Truelove condition, which enables a long-term integration. 
The polytropic index $\Gamma=2$ and transition density $n=3\times10^{18}\cm$ are adjusted to reproduce the protostellar  radius expected from recent calculations \citep{machida13a,tomida13}. 
With this equation of state, a second hydrostatic core (or protostar; \citealt{larson69}) $\sim8\rsun$ in size forms at $n \gtrsim  3\times 10^{18}\cm$ and almost maintains its initial radius for the duration of the calculation.  
It is worth noting that  the protostellar radius  ($\sim 8\rsun$) is twice as large as that in a one-dimensional calculation ($\sim4\rsun$; \citealt{masunaga00}) but is comparable to that in a recent three-dimensional calculation ($\sim10\rsun$; \citealt{tomida13}).
The numerical methods and protostellar model are fully described in our previous studies \citep{machida04, machida05a, machida05b,machida06,machida13b,machida14}.

\section{Results} 
\label{sec:results}
Starting from a prestellar stage, the cloud evolution was calculated until $\sim270$\,yr after the protostar formation.  
First, an overview of the cloud evolution calculated before (i.e., the gas collapsing phase) and just after (i.e., the early main accretion phase) the protostar formation is presented.
After the prestellar cloud begins to collapse, the first (hydrostatic) core \citep{larson69,masunaga00} appears when the central density reaches $n_c \sim10^{11}\cm$.
Subsequently, a low-velocity outflow is driven by the first core.
The first core gradually contracts and the second collapse begins at $n_c \gtrsim 10^{15}\cm$.
Finally, a protostar appears at $n_c \sim 5\times10^{18}\cm$ with a high-velocity jet emerging nearby. 
Following the protostar formation, the first core evolves into a rotation-supported circumstellar disk \citep{bate98,bate14,machida10a,tsukamoto13}. 

\begin{figure}
\includegraphics[width=150mm]{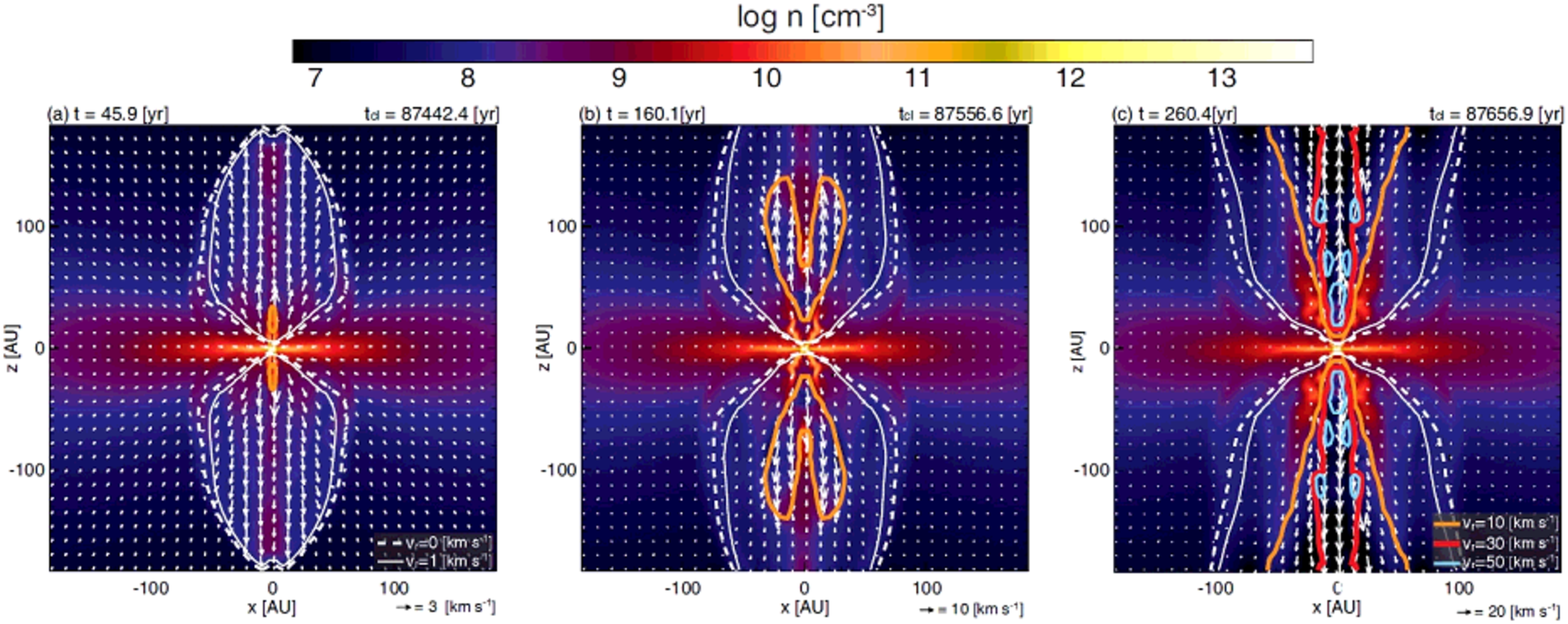}
\caption{ 
Cloud evolution on a large scale after protostar formation with a box size of $370$\,AU (see movie1). 
Density (color) and velocity (arrows; the velocity scale is given  in the lower side of each panel) distributions and the velocity contours (contour levels are shown in panels {\it a} and {\it c}) on the $y=0$ plane are plotted in each panel. 
The elapsed times since the protostar formation $t$ and since the start of the cloud collapsing $t_{\rm cl}$ are given in the upper side of each panel.  
}
\label{fig:1}
\end{figure}

In this calculation, the low-velocity outflow is driven by the first core prior to the protostar formation and by the circumstellar disk after the protostar formation. 
For convenience, I term the flow driven by the first core or large-scale disk the ``outflow'' and that driven near the protostar the ``jet''.
Figure~\ref{fig:1}{\it a} shows that the outflow extends up to $\sim 200$\,AU in $t=45.9$\,yr.
The low-velocity outflow depicted by the white contour has a velocity of $\lesssim 10\km$ and surrounds the high-velocity ($\gtrsim 10\km$) jet shown in orange.
The high-velocity jet driven near the protostar propagates into the low-velocity outflow as seen in Figures~\ref{fig:1}{\it b} and {\it c}; it reaches the head of the low-velocity outflow at $t=190$\,yr and thereafter precedes it. 
A  mass ejection, which is entrained by the high-velocity jet, can be observed just outside the orange contour in Figure~\ref{fig:1}{\it c}. 
The high-velocity jet transfers momentum to the low-velocity outflow, and the low-velocity flow gradually increases its speed.

Figure~\ref{fig:2}{\it a} shows that a relatively low-velocity jet ($\lesssim 50\km$; the white dotted contour) is driven by the disk surface near the protostar and has a wide opening angle. 
As shown in Figures~\ref{fig:2}{\it b} and {\it c}, a relatively high-velocity jet ($\gtrsim 50\km$; the orange contour) with good collimation appears just above and below the protostar and is embedded in the lower-velocity jet. 
After the higher-velocity jet transiently weakens (Fig.~\ref{fig:2}{\it d}), it is reactivated as seen in Figures~\ref{fig:2}{\it e} and {\it f}.  
Figure~\ref{fig:2} indicates that a relatively high-velocity jet ($\gtrsim 50\km$) with good collimation intermittently appears in the vicinity of the protostar, while a relatively low-velocity jet ($\lesssim 50\km$) with a wide opening angle is steadily driven by the disk near the protostar.
Therefore, the protostellar jet has a nested structure, in which higher velocity components are embedded in lower velocity components.

\begin{figure}
\includegraphics[width=150mm]{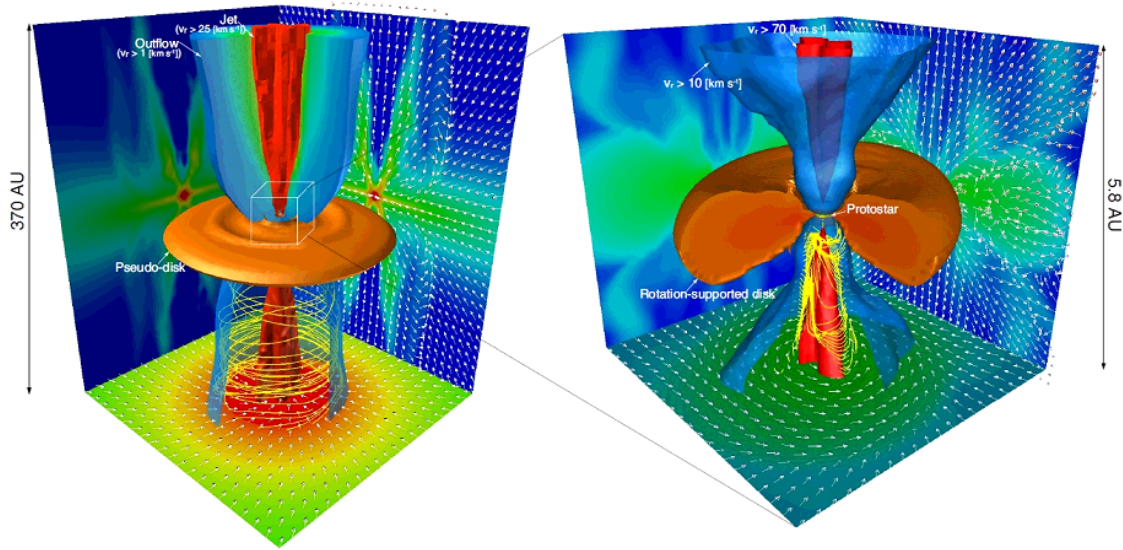}
\caption{
Three-dimensional structure at $t=266.9$\,yr on different scales (as specified next to the image, see movie3).
The right panel is a close-up view of box outlined in the left panel.
The low-velocity and high-velocity flows are depicted by blue and red volumes, respectively. 
The high density gas region, which corresponds to the pseudo disk (left) and rotation-supported disk (right), is plotted as an orange iso-density surface.
The density and velocity distributions on the $x=0$, $y=0$ and $z=0$ cutting planes are plotted on the corresponding wall surface. 
The magnetic field lines inside the $z<0$ region are shown by yellow streamlines. 
}
\label{fig:3}
\end{figure}

Figure~\ref{fig:3} also indicates that the high-velocity jet is enclosed by the low-velocity outflow (left panel) and has a nested velocity structure (right panel). 
In addition, the  high-velocity component is narrower than the low-velocity component.
The figure also shows  strongly twisted magnetic field lines within the outflow and jet.

\begin{figure}
\includegraphics[width=150mm]{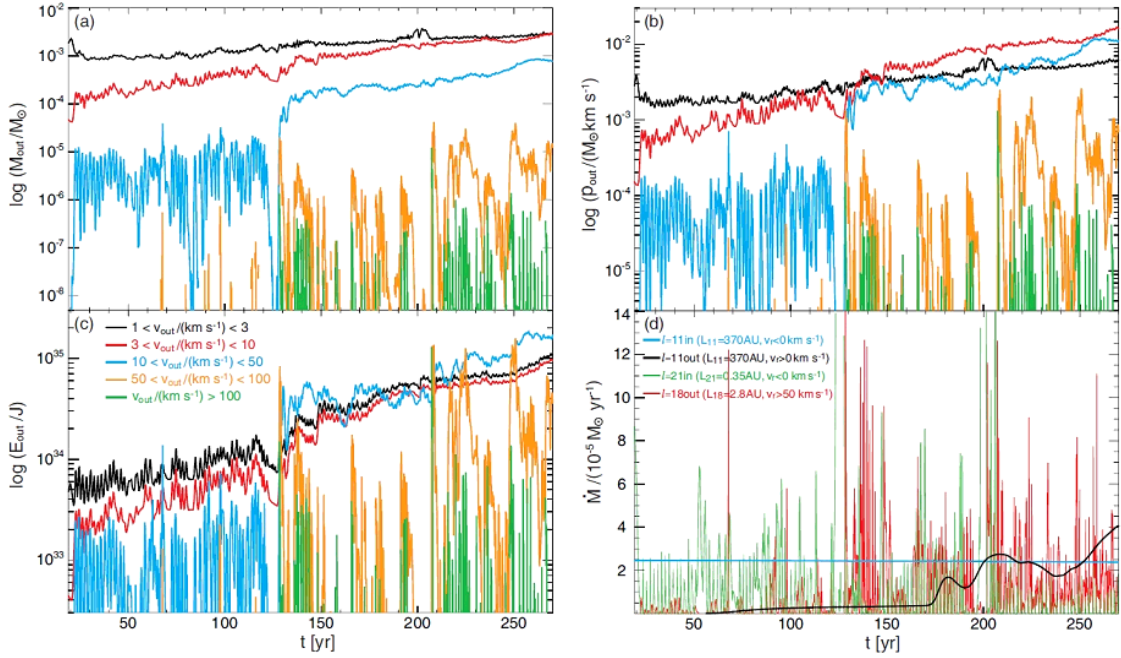}
\caption{ 
Panels ({\it a})-({\it c}): the mass ({\it a}), momentum ({\it b}) and kinetic energy ({\it c}) of the outflowing gas against the elapsed time $t$. 
The velocity range color-code for the outflowing gas is given in panel ({\it c}). 
Quantities are integrated over the velocity ranges given in panel ({\it c}).   
Panel ({\it d}): the accretion (blue and green) and outflow (black and red) rates against the elapsed time $t$. 
The inflow rate is estimated on the surface of the $l=11$ and $l=21$ grids with a box size of $L_{11}=370$\,AU and $L_{21}=0.35$\,AU, while the outflow rate is estimated on the $l=11$ and $l=18$ ($L_{18}=2.8$\,AU) grids. 
To extract only the high-velocity component, we integrated the outflowing gas for $v_r > 50\km$ to estimate the outflow rate on the $l=18$ grid. 
The grids ($l=18$ for inflow and 21 for outflow) used for estimating the inflow and outflow rates near the protostar were determined to eliminate the influence of gas circulation within the disk because the gas in the circumstellar disk sometimes moves outward with spiral arms.
}
\label{fig:4}
\end{figure}

Figure~\ref{fig:4}{\it a} shows that the mass of the low-velocity component ($1\km<v_{\rm out}<10\km$) is much larger than that of the high-velocity component ($v_{\rm out}>50\km$). 
The low-velocity outflow has a wide opening angle and can sweep up a large fraction of infalling gas. 
On the other hand, the high-velocity jet has a considerably narrower opening angle and  cannot incorporate a significant quantity of infalling gas. 
This indicates that the low-velocity outflow plays a more important role than the high-velocity jet in the early phase of the star formation.
The low-velocity wide-opening angle outflow seems to contribute to determine the star formation efficiency \citep{machida13a}.
However, a further long-term jet calculation is necessary,  because the jet may also eject a sufficient quantity of infalling gas into interstellar space.

The same trend is seen in the momentum of the outflowing gas (Fig.~\ref{fig:4}{\it b}), in which the momentum of the low-velocity component is larger than that of the high-velocity component. 
However, the flow with an intermediate velocity of $10\km < v_{\rm out}< 50\km$ (the blue line) has a comparable momentum to the lower velocity component of $v_{\rm out}< 10\km$ (the black and red lines). 
Thus, the high-velocity component (or the jet) is expected to partly contribute to the outflow propagation. 
Note that the low-velocity outflow is mainly driven by a large-scale disk and can extend to large distances without the support of the jet \citep{machida13a}. 
On the other hand, the kinetic energy of the jet sometimes exceeds that of the outflow as shown in Figure~\ref{fig:4}{\it c}. 
Figures~\ref{fig:4}{\it a}-{\it c} indicate that both the outflow and the jet can contribute to the dynamical evolution of the star-forming cloud. 

In Figures~\ref{fig:4}{\it a}-{\it c}, the properties of the high-velocity components vary greatly with time, indicating that the high-velocity jet appears intermittently as shown in Figure~\ref{fig:2}. 
To investigate the time variability, the inflow and outflow rates are plotted against the elapsed time in Figure~\ref{fig:4}{\it d}.
Since the protostar has a mass of $0.003\msun$ at its formation and $0.013\msun$ at the end of the calculation, the averaged accretion rate onto the protostar during the calculation is $\sim3.7\times 10^{-5}\msun$\,yr$^{-1}$.
The figure indicates that, on a large scale of 370\,AU ($l=11$), the inflow rate is almost constant at $\sim 2.5 \times 10^{-5} \mdot$ and the outflow rate gradually increases with time. 
On the other hand, the inflow and outflow rates exhibit strong time variability near the protostar (on a smaller scale).
The outflow rate is $\lesssim10^{-8}\mdot$ in a quiescent phase and exceeds $10^{-4}\mdot$ in an active phase on the $l=18$ grid. 
Around the protostar, the inflow rate shows a similar trend to the outflow rate.  
Figure~\ref{fig:4}{\it d} also indicates that, for $t \gtrsim$200\,yr, the outflow rate sometimes exceeds the inflow rate on the $l=11$ grid, because, in addition to the low-velocity outflow driven by the disk outer region, the jet contributes to the outflow rate in a large scale after the jet overtakes outflow. 

\section{Discussion}
As shown in \S\ref{sec:results}, the jet exhibits strong time-variability; however, it is difficult to determine the cause of this.
By contrast, non-steady accretion onto the protostar was shown in previous studies with a sink \citep[e.g.,][]{vorobyov06, machida11,machida13a}, in which the jet driving region was masked by sink cells.
Non-steady (episodic) accretion can be caused by gravitational instability \citep{vorobyov06}, a combination of gravitational instability and magnetic dissipation  \citep{machida11} or magnetorotational instability (MRI, \citealt{romanova12}). 
In addition,  magnetic braking plays a significant role in transporting angular momentum and promotes the (non-steady) gas accretion onto the protostar \citep{inutsuka10,seifried11,joos12,li13,machida14}.
Thus, the non-steady jet is considered to be a natural consequence of non-steady accretion.

\begin{figure}
\includegraphics[width=150mm]{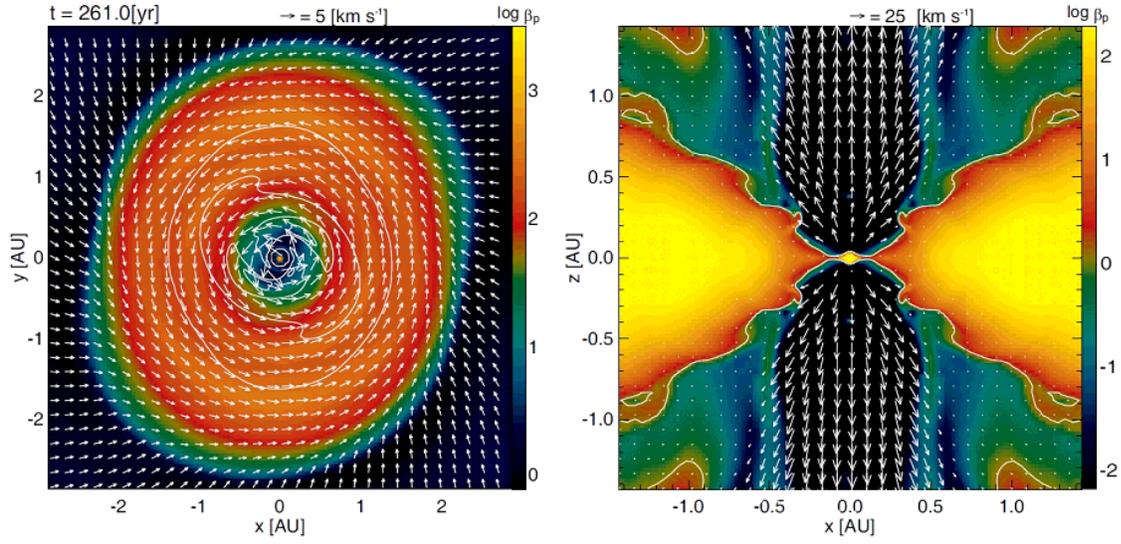}
\caption{
Plasma beta (color) and velocity (arrows) distributions at $t=261$\,yr on the $z=0$ (left) and $y=0$ (right) planes. 
The box scale differs between two panels. 
The white contour indicates the density distribution (left panel, $n=10^{14}$, $10^{15}$ and $10^{16}\cm$) and $\beta_p=10$ (right panel).
}
\label{fig:5}
\end{figure}

In our calculation, on the scale of the circumstellar disk ($r\sim1-3$\,AU), the angular momentum is mainly transported due to the effect of gravitational instability and magnetic braking. 
The angular momentum is transported by magnetic braking in the outer disk region ($r\gtrsim 2$\,AU) where the magnetic field is strong (plasma beta is low) and the magnetic dissipation is ineffective \citep{machida11b,li13}.
On the other hand, in the high-density disk region ($r\lesssim 2$\,AU), the magnetic field undergoes significant Ohmic dissipation \citep{nakano02} and the angular momentum is mainly transported by the non-axisymmetric structure caused by gravitational instability \citep{toomre64}. 
The left panel of Figure~\ref{fig:5} shows that a non-axisymmetric density pattern (contour) develops in the high plasma beta region (color) of  $0.5\,{\rm AU} \lesssim r \lesssim 2\, {\rm AU}$.

The gas density and temperature further increase near the protostar.
With a density of $n\gtrsim10^{15}\cm$ and temperature of $T\gtrsim1000$\,K, the thermal ionization of alkali metals increases the degree of ionization and reduces the efficiency of magnetic dissipation \citep{nakano02,machida07}.
As shown in Figure~\ref{fig:5}, the magnetic field is strong  (plasma beta is low) in the region of $r\lesssim 0.5$\,AU where the magnetic field is re-coupled with the gas. 
Thus, in this region, the rotational motion of the gas twists the magnetic field, which  in turn drives the high-velocity jet.
In the calculation, the jet has a variable time period $\sim1-2$\,yr from which the jet driving radius $r_c$ can be estimated assuming the Keplerian rotation to be 
\begin{equation}
r_c = \left( \dfrac{GM_{\rm ps}P^2}{4\pi^2} \right)^{1/3},
\label{eq:rc}
\end{equation}
where $M_{\rm ps}$ and $P$ are the protostellar mass and typical period of the jet variability, respectively. 
By substituting $P=2$\,yr and $M_{\rm ps}=0.013\msun$,  $r_c=0.37$\,AU is derived and  corresponds well to the jet driving region seen in the right panel of Figure~\ref{fig:5}.
On the other hand, a low-velocity outflow is driven by the outer disk region ($r\gtrsim2$\,AU) where the magnetic field is coupled with the gas due to a relatively high degree of ionization \citep{nakano02}.
The low-velocity outflow also contributes to the angular momentum transport on a large scale \citep{tomisaka00}.

The mechanism of angular momentum transport differs on different scales: the angular momentum is transported by magnetic braking and outflow ($r\gtrsim2$\,AU), gravitational instability ($0.5 {\rm AU} \lesssim r \lesssim 2$\,AU) and jet ($r \lesssim 0.5$\,AU).
In addition,  MRI may play a role in distributing the angular momentum on the surface of the disk around the protostar.
 
\begin{figure}
\includegraphics[width=150mm]{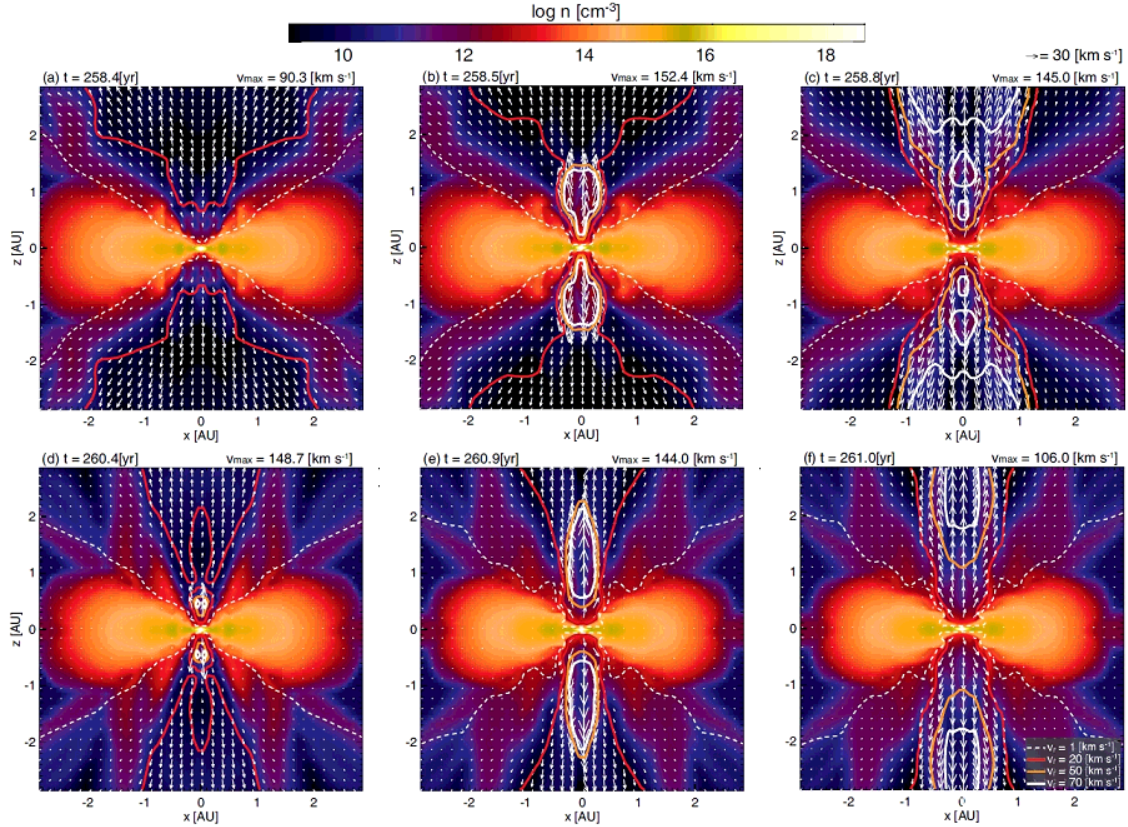}
\caption{
Cloud evolution around the protostar during the epochs of $t=258.4-261.0$\,yr with a box size of 5.8\,AU (see movie2). 
Density (color) and velocity (arrows, the velocity scale is given in the upper right) distributions on the $y=0$ plane are plotted in each panel. 
The velocity contours are also plotted (contour levels are described in panel {\it f}).
The elapsed time since the protostar formation $t$ and the maximum outflowing velocity $v_{\rm max}$ are also given for   each panel. 
}
\label{fig:2}
\end{figure}

MRI does not develop efficiently near the disk equatorial plane where the magnetic field significantly dissipates.
On the other hand, MRI can develop near the disk surface where magnetic dissipation is not effective, and mass accretion can occur from the disk surface layer \citep{fleming03}. 
In our calculation,  the disk surface layer has $\beta_{\rm p}\sim10$ and MRI can grow \citep{balbus91}.
Although a more detailed analysis is necessary, a wavy structure on the disk surface (Fig.~\ref{fig:2}) may be caused by MRI.
Note that recent studies indicate that at least 20 - 30 grid points per scale height are necessary to better resolve MRI turbulence \citep[e.g.,][]{hawley13}.
Thus, more spatial resolution may be necessary to calculate MRI: the scale height is resolved $\sim$ 10 -- 24 grid points in this study.
Moreover, the magnetic coupled region around the protostar is expected to extend with time \citep{inutsuka12}.
A higher spatial resolution and longer integration are necessary to further investigate the properties of the protostellar jet.

In this calculation, the jet reaches a maximum speed of $\gtrsim100\km$.
The Keplerian velocity $\Omega_{\rm ps}$ just outside  the protostar, however, is $\sim20\km$ with a protostellar mass of $0.013\msun$ and a radius of $\sim8\rsun$.
Thus, the maximum jet speed is several times higher than the Keplerian velocity.
Conversely, Figure~\ref{fig:5} right panel shows that the jet has a speed of $\sim20\km$ around the protostar.
Thus, the jet speed at its root corresponds to the Keplerian velocity.
Roughly, the jet terminal velocity can be obtained by $v_{\rm term} \sim \Omega_{\rm ps}\, r_{\rm A}$, where $r_{\rm A}$ is the Alfv\'en radius and estimated as $r_{\rm A}=0.1$\,AU. 
In the calculation, the jet terminal velocity is estimated to $v_{\rm term}\sim 40$\,km s$^{-1}$, which is somewhat slower than the maximum jet speed.
The jet speed increases with distance from the protostar and reaches a maximum of $\sim100\km$. 
Thus, it is likely that the jet is accelerated by mechanisms such as the Lorentz force \citep{kudoh97} and the Laval-nozzle as it propagates. 
The magnetic pressure gradient can also accelerate the jet \citep{spruit97,tomisaka02,machida08a,seifried12}. 
Although more in depth investigation is necessary to clarify the acceleration mechanism, this study indicates that a protostar can produce a high-velocity jet even in a very early phase of the star formation.

In this study, I reproduced a high-velocity jet with good collimation enclosed by a low-velocity outflow with  a wide opening angle.
Because the jet propagation was calculated for just $\sim270$\,yr, the jet's size is only $\sim400$\,AU.  
It may be difficult to observe small-sized jets embedded in the dense infalling envelope.
However, such compact jets are observable using ALMA. 

\acknowledgments
This work was supported by Grants-in-Aid from MEXT (25400232, 26103707).

\clearpage

\clearpage

\clearpage

\clearpage

\clearpage



\begin{thebibliography}{}
\bibitem[Arce \& Goodman(2002)]{arce02} 
Arce, H.~G., \& Goodman, A.~A.\ 2002, ApJ, 575, 928 

\bibitem[Arce et al.(2007)]{arce07} 
Arce, H.~G., Shepherd, D., Gueth, F., et al.\ 2007, Protostars and Planets V, 245 


\bibitem[Bate (1998)]{bate98} 
Bate, M.~R.\ 1998, ApJL, 508, L95


\bibitem[Bate et al.(2014)]{bate14} 
Bate, M.~R., Tricco, T.~S., \& Price, D.~J.\ 2014, MNRAS, 437, 77 

\bibitem[Balbus \& Hawley(1991)]{balbus91} 
Balbus, S.~A., \& Hawley, J.~F.\ 1991, \apj, 376, 214 

\bibitem[Fleming \& Stone(2003)]{fleming03} 
Fleming, T., \& Stone, J.~M.\ 2003, \apj, 585, 908 

\bibitem[Frank et al.(2014)]{frank14} 
Frank, A., Ray, T.~P., Cabrit, S., et al.\ 2014, arXiv:1402.3553 

\bibitem[Hawley et al.(2013)]{hawley13} 
Hawley, J.~F., Richers, S.~A., Guan, X., \& Krolik, J.~H.\ 2013, \apj, 772, 102 


\bibitem[Inutsuka et al.(2010)]{inutsuka10} 
Inutsuka, S., Machida, M.~N., \& Matsumoto, T.\ 2010, \apjl, 718, L58 

\bibitem[Inutsuka(2012)]{inutsuka12} 
Inutsuka, S.\ 2012, Progress of Theoretical and Experimental Physics, 2012, 010000 

\bibitem[Joos et al.(2012)]{joos12} 
Joos, M., Hennebelle, P., \& Ciardi, A.\ 2012, \aap, 543, A128 

\bibitem[Konigl \& Pudritz(2000)]{konigl00} 
Konigl, A., \& Pudritz, R.~E.\ 2000, Protostars and Planets IV, 759 

\bibitem[Kudoh \& Shibata(1997)]{kudoh97} 
Kudoh, T., \& Shibata, K.\ 1997, \apj, 474, 362 

\bibitem[Larson(1969)]{larson69} 
Larson, R. B., 1969, MNRAS, 145, 271.

\bibitem[Li et al.(2013)]{li13} 
Li, Z.-Y., Krasnopolsky, R., \& Shang, H.\ 2013, \apj, 774, 82 

\bibitem[Machida et al.(2004)]{machida04} 
Machida, M. N., Tomisaka, K., \& Matsumoto, T.\ 2004, MNRAS, 348, L1 

\bibitem[Machida \etal(2005a)]{machida05a}
Machida, M. N., Matsumoto, T., Tomisaka, K., \& Hanawa, T. 2005, MNRAS, 362, 369

\bibitem[Machida \etal(2005b)]{machida05b} 
Machida, M. N., Matsumoto, T., Hanawa, T., \& Tomisaka, K. 2005b, MNRAS, 362, 382 

\bibitem[Machida et al.(2006)]{machida06} 
Machida, M.~N., Matsumoto, T., Hanawa, T., \& Tomisaka, K.\ 2006, ApJ, 645, 1227 

\bibitem[Machida et al.(2007)]{machida07} 
Machida, M.~N., Inutsuka, S., \& Matsumoto, T.\ 2007, \apj, 670, 1198 

\bibitem[Machida et al.(2008a)]{machida08a} 
Machida, M.~N., Inutsuka, S., \& Matsumoto, T.\ 2008a, \apj, 676, 1088 

\bibitem[Machida et al.(2008b)]{machida08b} 
Machida, M.~N., Tomisaka, K., Matsumoto, T., \& Inutsuka, S.\ 2008b, \apj, 677, 327 

\bibitem[Machida et al.(2010a)]{machida10a} 
Machida, M.~N., Inutsuka, S., \& Matsumoto, T.\ 2010a, ApJ, 724, 1006 

\bibitem[Machida et al.(2011a)]{machida11} 
Machida, M.~N., Inutsuka, S., \& Matsumoto, T.\ 2011a, \apj, 729, 42 

\bibitem[Machida et al.(2011b)]{machida11b} 
Machida, M.~N., Inutsuka, S., \& Matsumoto, T.\ 2011b, \pasj, 63, 555 

\bibitem[Machida \& Matsumoto(2012)]{machida12} 
Machida, M.~N., \& Matsumoto, T.\ 2012, \mnras, 421, 588 

\bibitem[Machida \& Hosokawa(2013a)]{machida13a} 
Machida, M.~N., \& Hosokawa, T.\ 2013a, \mnras, 431, 1719 

\bibitem[Machida \& Doi(2013b)]{machida13b} 
Machida, M.~N., \& Doi, K.\ 2013b, \mnras, 435, 3283 

\bibitem[Machida et al.(2014)]{machida14} 
Machida, M.~N., Inutsuka, S., \& Matsumoto, T.\ 2014, \mnras, 438, 2278 

\bibitem[Masunaga \& Inutsuka(2000)]{masunaga00} 
Masunaga, H., \& Inutsuka, S., 2000, ApJ, 531, 350

\bibitem[Matzner \& McKee(2000)]{matzner00} 
Matzner, C.~D., \& McKee, C.~F.\ 2000, \apj, 545, 364 


\bibitem[Mouschovias \& Spitzer(1976)]{mouschovias76} 
Mouschovias, T.~C., \& Spitzer, L., Jr.\ 1976, \apj, 210, 326 

\bibitem[Mundt \& Fried(1983)]{mundt83} 
Mundt, R., \& Fried, J.~W.\ 1983, ApJL, 274, L83 

\bibitem[Nakano et al.(1995)]{nakano95} 
Nakano, T., Hasegawa, T., \& Norman, C.\ 1995, \apj, 450, 183 

\bibitem[Nakano et al.(2002)]{nakano02} 
Nakano, T., Nishi, R., \& Umebayashi, T.\ 2002, \apj, 573, 199 

\bibitem[Pudritz et al.(2007)]{pudritz07} 
Pudritz, R.~E., Ouyed, R., Fendt, C., \& Brandenburg, A.\ 2007, Protostars and Planets V, 277 

\bibitem[Ren et al.(2011)]{ren11} 
Ren, J.~Z., Liu, T., Wu, Y., \& Li, L.\ 2011, \mnras, 415, L49 


\bibitem[Romanova et al.(2012)]{romanova12} 
Romanova, M.~M., Ustyugova, G.~V., Koldoba, A.~V., \& Lovelace, R.~V.~E.\ 2012, \mnras, 421, 63 

\bibitem[Seifried et al.(2011)]{seifried11} 
Seifried, D., Banerjee, R., Klessen, R.~S., Duffin, D., \& Pudritz, R.~E.\ 2011, \mnras, 417, 1054 

\bibitem[Seifried et al.(2012)]{seifried12} 
Seifried, D., Pudritz, R.~E., Banerjee, R., Duffin, D., \& Klessen, R.~S.\ 2012, \mnras, 422, 347 

\bibitem[Spruit et al.(1997)]{spruit97} 
Spruit, H.~C., Foglizzo, T., \& Stehle, R.\ 1997, \mnras, 288, 333 

\bibitem[Tomida et al.(2013)]{tomida13} 
Tomida, K., Tomisaka, K., Matsumoto, T., et al.\ 2013, ApJ, 763, 6 

\bibitem[Tomisaka(1998)]{tomisaka98} 
Tomisaka, K.\ 1998, \apjl, 502, L163 

\bibitem[Tomisaka(2000)]{tomisaka00} 
Tomisaka, K.\ 2000, \apjl, 528, L41 

\bibitem[Tomisaka(2002)]{tomisaka02} 
Tomisaka, K.\ 2002, ApJ, 575, 306 

\bibitem[Toomre(1964)]{toomre64} 
Toomre, A.\ 1964, \apj, 139, 1217 

\bibitem[Tsukamoto et al.(2013)]{tsukamoto13} 
Tsukamoto, Y., Machida, M.~N., \& Inutsuka, S.\ 2013, MNRAS, 436, 1667 

\bibitem[Truelove et al.(1997)]{truelove97}
Truelove J, K., Klein R. I., McKee C. F., Holliman J. H., Howell L. H., \& Greenough J. A., 1997, ApJ, 489, L179

\bibitem[Velusamy et al.(2007)]{velusamy07} 
Velusamy, T., Langer, W.~D., \& Marsh, K.~A.\ 2007, ApJL, 668, L159 

\bibitem[Vorobyov \& Basu(2006)]{vorobyov06} 
Vorobyov, E.~I., \& Basu, S.\ 2006, \apj, 650, 956 


\end{thebibliography}
\end{document}